\documentstyle[epsfig]{elsart}

\begin{document}

\begin{frontmatter}
 
\vspace*{2cm}
\title{Simulation studies of the HADES first level trigger.\\
PART II: Performance in hadron induced reactions}

\author{R.~Schicker\thanksref{corr}}
\author{and H.~Tsertos}
\address{University of Cyprus, Nicosia, Cyprus}

\begin{abstract}

{\normalsize
The HADES first level trigger is studied for the system p+Ni at a beam energy 
of 2 AGeV. The timing properties of the trigger signal are reported. The 
efficiency loss due to deadtime is specified. A trigger requirement of a time 
overlap window with the start detector is described. The trigger rates for 
different overlap windows are given.
}

\end{abstract}

\thanks[corr]{Corresponding author, e-mail "schicker@alpha2.ns.ucy.ac.cy" \\
Dept. Nat. Science, Univ. Cyprus, PO 537, 1678 Nicosia, Cyprus}

\end{frontmatter}

\vspace{-12.5cm}
\hspace{10.cm}{\bf UCY--PHY--96/15}
\vspace{11.5cm}

\newpage

\section{Introduction} 
\label{sec:intro}

The dilepton spectrometer HADES is currently being built at the heavy-ion 
synchrotron SIS at GSI Darmstadt\cite{hades1}. HADES will measure dielectron 
pairs emitted in proton-proton, proton-nucleus and nucleus-nucleus collisions 
in the beam energy range of 1-2 AGeV. Additionally, a secondary pion beam 
facility of momenta between 0.5 GeV/c and 2.5 GeV/c will allow the measurements 
of dilepton observables in pion induced reactions\cite{pion}. 

The reconstruction of dileptons from in-medium decays of $\rho,\omega$ and 
$\phi$ mesons allows to test conjectures about in-medium behavior 
of these mesons\cite{Brown}. Model calculations predict a change of the 
in-medium meson mass as a function of the nuclear density reached in the 
collision. Hence, proton and pion induced reactions yield information at 
ground state density, whereas heavy-ion induced collisions do so at 
densities up to three times the ground state density.

HADES is designed to operate at proton beam intensities of 10$^{8}$ per second. 
A 1\% interaction target results in 10$^{6}$ minimum bias events per second. 
The maximum pion beam intensity, on the other hand, reaches values of about 
2$\times$10$^{7}$ per spill at a momentum of 1 GeV/c. The proton beam is 
therefore more demanding on the first level trigger, and simulation results 
are shown below only for proton induced reactions. 

The purpose of this paper is to present simulation studies of the HADES first 
level trigger in hadron induced reactions. Simulation results are shown for 
the system p+Ni at 2 AGeV, in order to illustrate the trigger performance in 
a hadron induced very low multiplicity collision system.

This paper is organized as follows: Section \ref{sec:ftrig} gives a summary 
on the HADES first level trigger requirements in hadron induced reactions. 
In Section \ref{sec:pdat}, the production and analysis of the first level 
trigger simulation data are described. In Section \ref{sec:fpni}, the 
performance characteristics of the first level trigger in the very low 
multiplicity system p+Ni are presented. 

\section{First level trigger requirements}
\label{sec:ftrig}

In hadron induced reactions, the HADES first level trigger has to provide 
a reaction trigger. This reaction trigger can be derived from 
the multiplicity of the highly segmented time of flight (TOF) array. 
A condition on the number of TOF paddles carrying coincident signals 
will provide this trigger. 

Depending on the specific reaction channel to be measured, the total
multiplicity condition imposed on the TOF paddles can be as low as two. 
This multiplicity value corresponds to the two charged leptons of the pair. 
To illustrate this point, we mention here the p($\pi^{-}$,e$^{+}$e$^{-}$)n 
reaction which is tagged by a multiplicity condition of two. This reaction 
attracts considerable interest for the investigation of the time-like nucleon 
form factor below the threshold accessible in nucleon-antinucleon annihilation. 
Thus, a versatile first level trigger for hadron induced reactions is 
characterized by a multiplicity condition of two or larger.

The trigger signal derived from the multiplicity condition of the TOF paddles 
is used as gate for the ADCs of the RICH detector and as common STOP for the 
TDCs. Thus, the delay of this signal with respect to the time of reaction as 
well as the time jitter are of particular interest. The time jitter of the first
level trigger signal arises from different sources. First, trajectory length 
variations over the polar angular range of the spectrometer induce particle 
TOF variations. Second, velocity variations of the particles defining the 
trigger add to the TOF variations. Third, the different signal propagations 
in the TOF paddles, depending on the location of the hit point, add varying 
delays to the TOF signals.

A multiplicity condition of two is triggered by the two fastest particles of the 
event, i.e., by the two leptons ($\gamma \geq 20$) of the pair. The contribution 
to the time jitter due to particle velocity variations mentioned above therefore
vanishes. Thus, for a multiplicity condition of two, the trigger time jitter of 
dilepton events is considerably reduced as compared to events without dileptons. 
This time correlation allows an additional timing condition for the trigger 
signal derived from the fastest two particles of the reaction. Events containing
a dilepton will meet this condition, but most of the events without dileptons 
will not. Hence, this requirement will considerably improve the deadtime of 
the first level trigger system. Such a timing condition necessitates, however, 
an independent measurement of the reaction time by another detector system.

The first level trigger performance in the p+Ni system depends weakly on the 
duration of the TOF paddle signals\cite{UCY3}. In this report, all the results 
shown have been derived with a TOF signal length of 15 nsec.

\section{First level trigger data production and analysis}
\label{sec:pdat}

As in the earlier investigations, the full HADES geometry was implemented into 
the GEANT package\cite{LVL1HI}. A realistic field map of the toroidal 
magnetic field is used for tracking of the charged particles\cite{Heike}.

The collisions of the p+Ni system are modeled by a transport equation of the 
Boltzmann-Uehling-Uhlenbeck (BUU) type. The dynamical evolution of the 
collisions is determined by calculating the phase space evolution for nucleons, 
Delta and N$^{*}$ resonances. With this code, good agreement is found between 
data and model predictions in proton induced collisions in the energy range 
1-2 AGeV\cite{Wolf}. 

In the system p+Ni presented here, the production and analysis of the first 
level trigger data proceed in a similar manner as for the heavy-ion systems 
reported \cite{LVL1HI}.

\section{First level trigger in p+Ni collisions}
\label{sec:fpni}

\subsection{Minimum total multiplicity}

Due to the very low multiplicity of the p+Ni system, a large fraction of 
reactions has no tracks in one or more of the azimuthal sectors. As in the 
Ne+Ne system studied earlier\cite{LVL1HI}, the minimal total multiplicity 
condition M$_{L}$ is therefore used in order to define the trigger. 
For all of the results shown in this report, a multiplicity condition
M$_{L} \geq 2$ is used. Thus, the two fastest particles of an event define
the trigger.   

\subsection{Trigger timing}

Fig. \ref{fig:fig1} shows the trigger timing of p+Ni events with different impact 
parameters. Here, the trigger is defined by the condition of minimum total 
multiplicity M$_{L} \geq 2$. The time zero is the time of reaction. Events 
with impact parameter b\,=\,1\,fm are represented by the solid line. The 
FWHM of their time distribution amounts to about 10 nsec. Events with impact 
parameters of 2 and 3\,fm  exhibit a very similar behavior. The distributions 
in Fig. \ref{fig:fig1} have considerable tails for high $\Delta$T$_{0}$ values. 
These tails result from events in which the second particle of the trigger 
condition is very slow.

\subsection{Trigger timing of events containing e$^{+}$e$^{-}$ pairs}

The width of the trigger timing shown in Fig. \ref{fig:fig1} results from three 
different sources as discussed in Section \ref{sec:ftrig}. For reactions 
containing dileptons, the trigger timing derived from the two fastest particles,
the two leptons, does not include the jitter contribution due to particle 
velocity variations. Thus, the timing of the trigger signal is expected to be 
considerably improved as compared to events without dileptons. The dashed
line in Fig. \ref{fig:fig2} shows the timing of the trigger derived from 
the M$_{L} \geq 2$ condition for p+Ni events which do not contain dileptons. 
The solid line in Fig. \ref{fig:fig2} shows the trigger timing if the 
fastest two particles of the event are leptons of momenta larger than 100 MeV/c.
The width of this distribution is about 3 nsec. In Fig. \ref{fig:fig2}, the 
events without dileptons are downscaled with respect to the dilepton events. 
The relative intensity of these two classes of events is therefore arbitrary.

The width of the trigger timing of e$^{+}$e$^{-}$ events shown in Fig.
\ref{fig:fig2} results from variations in signal propagation delays in the 
TOF paddles and from trajectory length variations over the polar angle of 
the spectrometer. However, this width can be further reduced by using a mean 
timer circuit which provides TOF paddle timing independent of particle hit 
location. Additionally, the mean timer TOF signal can be corrected for 
the average trajectory length of the different paddles by introducing
appropriate cable delays. Only second order length variations within one TOF 
paddle of the two oppositely charged leptons remain in such a scheme. 

The improved timing of e$^{+}$e$^{-}$ events allows to reject triggers which do 
not arrive within a time window  $\Delta$T$_{0}$ following the reaction. This 
additional trigger condition is, however, contingent upon an independent 
measurement of the time of reaction with an accuracy comparable or better than 
the 3 nsec achieved by the TOF paddles. This independent measurement of the 
time of reaction can, for example, be achieved by the HADES start detector. A 
suitable time overlap coincidence between the HADES start detector and the 
trigger signal from the TOF paddles rejects a large fraction of events without 
dileptons. These rejected events do not generate a trigger, and the deadtime of 
the first level trigger system is therefore significantly improved.

In this report, deadtime losses are treated in the same way as in the
heavy-ion systems studied earlier\cite{LVL1HI}. A factor R$_{DT}$ is defined
which contains the losses due to the system deadtime. Thus, the first level
trigger rate is given by the reaction rate multiplied by the product of
trigger efficiency times R$_{DT}$.

Table \ref{tab:tab1} lists trigger rates and R$_{DT}$ values derived from the 
multiplicity condition M$_{L} \geq 2$. These rates are shown for deadtimes 
T$_{0}$ = 0,6 and 10 $\mu$sec and for time windows of $\Delta$T$_{0}$\,=\,12,16
and 60 nsec. Here, the time window $\Delta$T$_{0}$ represents the time interval
after the time of reaction during which the trigger signal has to arrive in 
order to generate a trigger. The condition $\Delta$T$_{0}$\,=\,60\,nsec 
represents a wide open time window and is therefore equivalent to no time 
condition. For each deadtime, the rate is shown in the 
left column in units of 10$^{5}$ per second. The R$_{DT}$ value is displayed 
in the corresponding column on the right. The improvement of the first level 
trigger performance with narrower time windows $\Delta$T$_{0}$ is clearly seen 
in Table \ref{tab:tab1}. For the expected deadtime of 10 $\mu$sec, a time 
window condition $\Delta$T$_{0}$ = 12\,nsec increases the number of dilepton 
events passed to the next trigger stage by a factor larger than three. This 
factor assumes, however, that the losses of dilepton events due to applying 
this time window condition $\Delta$T$_{0}$ can be neglected. As indicated in 
Fig. \ref{fig:fig2} by the vertical dotted line, a $\Delta$T$_{0}$\,=\,12\,nsec 
cut eliminates about 10\% of the dilepton events in the present simulations. 
It is, however, anticipated that the timing of the dilepton events can be 
further improved as discussed above. Hence, the present loss of 10\% dilepton
events will be reduced to a negligible level. The information shown in Table 
\ref{tab:tab1} represents therefore realistic expectations for the HADES 
first level trigger performance in the very low multiplicity system p+Ni.

\section{Conclusions}

Simulations of the HADES first level trigger in the p+Ni system indicate 
that a reaction trigger can be implemented by a multiplicity condition
imposed on the TOF paddles. A significant improvement of the first level
trigger performance is possible by making use of the timing derived from
the fastest two particles of the event. The achieved suppression of events
not containing dileptons results in a much reduced system deadtime. The 
number of dilepton events passed to the next trigger stage is correspondingly 
increased by a factor larger than three. The first level trigger of the 
p+Ni system resulting from such an architecture fulfills the rate requirement 
of 10$^{5}$ events per second.  

\begin{ack}

The support of the lepton group at GSI and, in particular, fruitful 
discussions with W.Koenig are gratefully acknowledged. The authors thank 
Gy.Wolf for providing the BUU data files used in the simulations.

\end{ack}

\newpage
{\Large \bf
\noindent
Figure Captions
}

\begin{figure}[h]
\vspace{.5cm}
\caption{                 
Timing of first level trigger from condition M$_{L} \geq 2$ for impact 
parameters of b=1,2 and 3\,fm. The time zero is the time of reaction.
}
\label{fig:fig1}
\end{figure}

\begin{figure}[h]
\vspace{.5cm}
\caption{                 
Timing of first level trigger from condition M$_{L} \geq 2$ for p+Ni events 
with dileptons (solid line) and for events without dileptons (dashed line). 
The vertical dotted line represents a time window condition 
$\Delta$T$_{0}$\,=\,12\,nsec (see text). The time zero is the time of reaction. 
}
\label{fig:fig2}
\end{figure}

\vspace{1.5cm}
{\Large \bf
\noindent
Tables
}

\vspace{.5cm}

\begin{table}[h]
\begin{tabular}{||c||c|c||c|c||c|c||} \hline
p+Ni&\multicolumn{2}{c||}{T$_{0}$\,=\,0\,$\mu$sec }
&\multicolumn{2}{c||}{T$_{0}$\,=\,6\,$\mu$sec }
&\multicolumn{2}{c||}{T$_{0}$\,=\,10\,$\mu$sec } \\ \cline{2-7}
&rate[10$^5$]&R$_{DT}$&rate[10$^5$]&R$_{DT}$&rate[10$^5$]&R$_{DT}$\\ \cline{1-7}
$\Delta$T$_{0}$\,=\,60\,nsec& 4.99&1.0 & 1.27&.25 & .840&.17 \\ \hline
$\Delta$T$_{0}$\,=\,16\,nsec& 2.03&1.0 & .926&.46 & .676&.33 \\ \hline
$\Delta$T$_{0}$\,=\,12\,nsec& .771&1.0 & .529&.69 & .435&.56 \\ \hline

\end{tabular}
\vspace{.2cm}
\caption{
First level trigger rates from condition M$_{L} \geq 2$ in the system p+Ni 
for deadtimes T$_{0}$ = 0,6 and 10\,$\mu$sec and for window conditions 
$\Delta$T$_{0}$ = 12,16 and 60 nsec. In the left column, the rates are shown
in units of 10$^{5}$ per second. In the right column, the deadtime 
reduction factor R$_{DT}$ is shown. 
\label{tab:tab1}
}
\end{table}

\pagestyle{empty}

\newpage

\begin{minipage}{15.cm}
\epsfig{figure=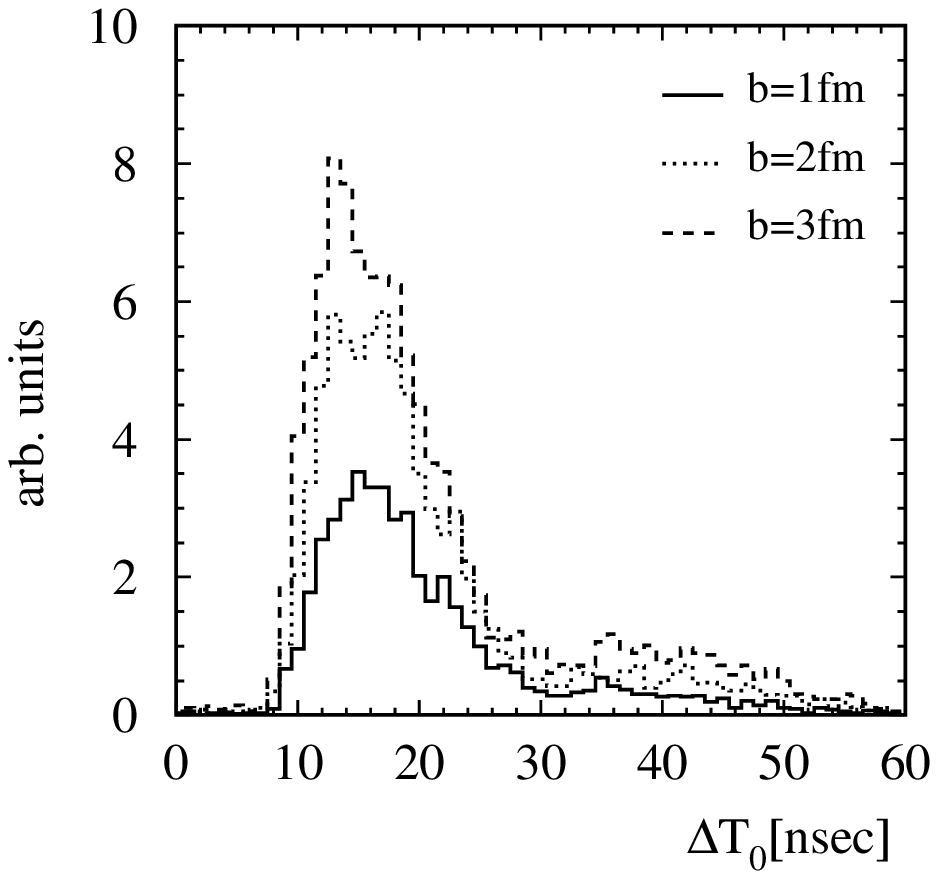,width=.95\linewidth}
\hfill
\end{minipage}

\begin{minipage}{2.cm}
\vspace{3.cm}
{\bf\huge FIG.1}
\end{minipage}

\newpage

\begin{minipage}{15.cm}
\epsfig{figure=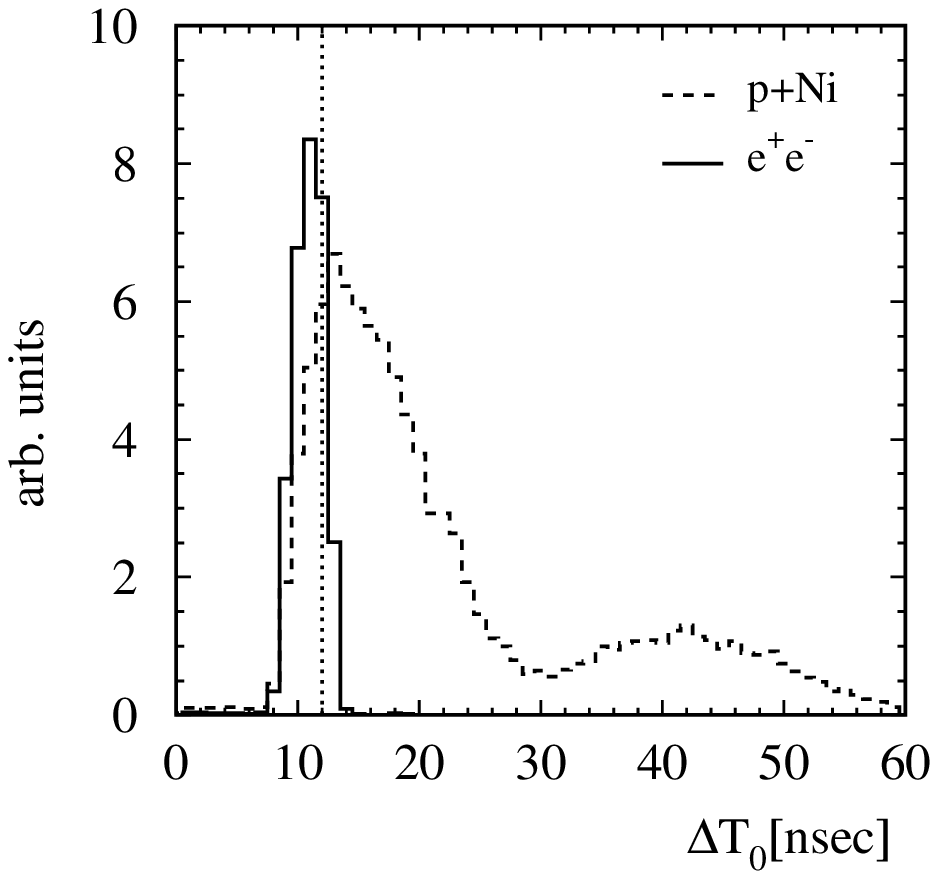,width=.95\linewidth}
\hfill
\end{minipage}

\begin{minipage}{2.cm}
\vspace{3.cm}
{\bf\huge FIG.2}
\end{minipage}

\end{document}